\documentclass[longauth]{aa} 
\usepackage{graphicx}
\usepackage{txfonts}
\usepackage{lscape}    
\usepackage{ulem}

\begin{document}

   \title{The Gaia-ESO survey: Lithium abundances in open cluster Red Clump stars \thanks{Based on observations collected with the FLAMES instrument at
VLT/UT2 telescope (Paranal Observatory, ESO, Chile), for the Gaia-
ESO Large Public Spectroscopic Survey (188.B-3002, 193.B-0936, 197.B-1074).}}
\titlerunning{Li abundance in red clump stars} 
\authorrunning{Magrini et al.}
\author{L. Magrini\inst{\ref{oaa}} \and 
R. Smiljanic\inst{\ref{ncac}} \and 
E. Franciosini\inst{\ref{oaa}} \and  
L. Pasquini\inst{\ref{eso}} \and 
S. Randich\inst{\ref{oaa}} \and 
G. Casali\inst{\ref{oaa},\ref{unifi}}\and 
C. Viscasillas V{\'a}zquez\inst{\ref{vilnius}}\and 
A. Bragaglia\inst{\ref{oabo}}\and 
L. Spina\inst{\ref{oapd}}\and 
K. Biazzo\inst{\ref{oarm}}\and 
G. Tautvai{\v s}ien{\. e}\inst{\ref{vilnius}} \and 
T. Masseron\inst{\ref{iac},\ref{ulaguna}} \and
M. Van der Swaelmen\inst{\ref{oaa}},
E. Pancino\inst{\ref{oaa},\ref{ssdc}}\and 
F. Jim\'enez-Esteban\inst{\ref{dipmadrid}}\and 
G. Guiglion\inst{\ref{leibniz}}\and  
S. Martell\inst{\ref{sidney}}\and  
T. Bensby\inst{\ref{lund}}\and 
V. D'Orazi\inst{\ref{oapd}}\and 
M. Baratella\inst{\ref{oapd}, \ref{unipd}}\and 
A. Korn\inst{\ref{lund}}\and 
P. Jofre\inst{\ref{dpchile}}\and
G. Gilmore\inst{\ref{cambridge}}\and
C. Worley\inst{\ref{cambridge}}\and
A. Hourihane\inst{\ref{cambridge}}\and
A. Gonneau\inst{\ref{cambridge}}\and
G.~G. Sacco\inst{\ref{oaa}} \and 
L. Morbidelli\inst{\ref{oaa}} }

\institute{INAF - Osservatorio Astrofisico di Arcetri, Largo E. Fermi 5, 50125, Firenze, Italy \email{laura.magrini@inaf.it} \label{oaa} 
\and
Nicolaus Copernicus Astronomical Center, Polish Academy of Sciences, ul. Bartycka 18, 00-716, Warsaw, Poland\label{ncac} 
\and
ESO, Karl Schwarzschild Strasse 2, 85748 Garching, Germany \label{eso} 
\and
Dipartimento di Fisica e Astronomia, Universit\`a degli Studi di Firenze, via G. Sansone 1, 50019 Sesto Fiorentino (Firenze), Italy\label{unifi} 
\and
Institute of Theoretical Physics and Astronomy, Vilnius University, Sauletekio av. 3, 10257 Vilnius, Lithuania \label{vilnius} 
\and
INAF - Osservatorio di Astrofisica e Scienza dello Spazio di Bologna, via Gobetti 93/3, 40129, Bologna, Italy\label{oabo} 
\and
INAF - Padova Observatory, Vicolo dell'Osservatorio 5, 35122 Padova, Italy\label{oapd} 
\and
INAF - Rome Observatory, Via Frascati, 33, Monte Porzio Catone (RM), Italy\label{oarm} 
\and
Instituto de Astrof\'{\i}sica de Canarias, E-38205 La Laguna, Tenerife, Spain\label{iac} 
\and
Universidad de La Laguna, Dept. Astrof\'{\i}sica, E-38206 La Laguna, Tenerife, Spain \label{ulaguna} 
\and
Space Science Data Center - Agenzia Spaziale Italiana, via del Politecnico, s.n.c., I-00133, Roma, Italy\label{ssdc}
\and
Departamento de Astrof\'{\i}sica, Centro de Astrobiolog\'{\i}a (CSIC-INTA), ESAC Campus, Camino Bajo del Castillo s/n, E-28692 Villanueva de la Ca\~nada, Madrid, Spain\label{dipmadrid}
\and
Leibniz-Institut f\"ur Astrophysik Potsdam (AIP) An der Sternwarte 16, 14482 Potsdam \label{leibniz}
\and
School of Physics, University of New South Wales, Sydney, NSW 2052, Australia\label{sidney}
\and
Lund Observatory, Department of Astronomy and Theoretical Physics, Box 43, SE-221 00 Lund, Sweden\label{lund}
\and
Dipartimento di Fisica e Astronomia {\it Galileo Galilei}, Vicolo Osservatorio 3, I-35122, Padova, Italy\label{unipd}
\and
N\'ucleo de Astronom\'ia, Facultad de Ingenier\'ia y Ciencias, Universidad Diego Portales, Ej\'ercito 441, Santiago, Chile\label{dpchile}
\and
Institute of Astronomy, University of Cambridge, Madingley Road, Cambridge CB3 0HA, United Kingdom\label{cambridge}
}
   \date{}

 
  \abstract
   {It has recently been suggested that all giant stars with mass below 2 $M_{\odot}$ suffer an episode of surface lithium enrichment between the tip of the red giant branch (RGB) and the red clump (RC).}
   {We test if the above result can be confirmed in a sample of RC and RGB stars that are members of open clusters.}
   {We discuss Li abundances in six open clusters with ages between 1.5 and 4.9~Gyr (turn-off masses between 1.1 and 1.7~$M_{\odot}$). These observations are compared with the predictions of different models that include rotation-induced mixing, thermohaline instability, mixing  induced by the first He flash, and energy losses by neutrino magnetic moment.}
   { In  six clusters, we find about 35\% RC stars with Li abundances that are similar or higher than those of upper RGB stars. This can be a sign of fresh Li production. Because of the extra-mixing episode connected to the luminosity bump, the expectation was for RC stars to have systematically lower surface Li abundances. However, we cannot confirm that the possible Li production is ubiquitous. For about 65\% RC giants we can only determine abundance upper limits that could be hiding very low Li abundances.}
   {Our results indicate a possible production of Li during the RC, at levels that would not classify the stars as Li rich. Determination of their carbon isotopic ratio would help to confirm that the RC giants have suffered extra mixing followed by Li enrichment. The Li abundances of the RC stars can be qualitatively explained by the models with an additional mixing episode close to the He flash.}


   \maketitle
%

\section{Introduction}
The existence of lithium-rich giants has been known for many years \citep[see, e.g.][]{brown89,  cb00, monaco11, kumar11}. However, their nature is a mystery that has not found a definitive solution yet. Many studies have shown that they amount to $\sim$1--2\% of all giant stars \citep[see, e.g.][]{casey16, smi18, deeppak19, martell20, c20}. Recent results, based on a combination of spectroscopic and asteroseismic observations, are showing the predominance of Li-rich giants in the core-helium-burning ‘red clump’ (RC) phase \citep{2014ApJ...784L..16S, casey19,kumar20a, 2021MNRAS.tmp.1175D,yan21,singh19, sing21}.
\citet{kumar20b} performed a large-scale investigation of the Li content in field stars at the RC phase and claimed that most of them show high levels of surface Li abundance for their evolutionary stage, with A(Li){\footnote{A(Li)=log($\frac{X(Li)}{X(H)}\cdot \frac{A_{\rm H}}{A_{\rm Li}}$)+12, where X and A are the mass fraction and the atomic mass}}$\,>-$0.9. They thus suggested that there is a systematic production of Li in low-mass stars (i.e., those with mass below 2 M$_{\odot}$) between the tip of the red giant branch (RGB) and the RC. 
The Li abundance in cluster giants were originally discussed by \citet{pasquini01}, who presumed that RC stars could have more Li than stars on the first ascent of the RGB. Subsequent observations  of giant stars in the open cluster IC~4651 \citep{pasquini04} supported the hypothesis, since they observed only upper limits for the stars that are found away from the RC, while they could measure Li in the RC stars.

Here, we exploit the sixth internal data release of the Gaia-ESO Survey \citep[{\sc idr6},][]{Gil,randich13} to explore the problem of Li abundances in RC stars from the point of view of open clusters. These objects give us the possibility 
to determine ages and main-sequence turn-off (MSTO) masses with high accuracy from isochrone fitting of their evolutionary sequence. The MSTO masses can be further used as proxy of the masses of the evolved stars. The comparison with isochrones also allows us to separate the RC stars from the RGB stars. We can thus study the evolution of the Li surface abundance along stellar evolutionary phases in different mass ranges. 

\section{Sample selection}
\label{section:abundance_and_sample}

For our analysis we use a selection of member stars from the sample of 57 open clusters with stellar parameters from
Gaia-ESO {\sc idr6} and analyzed in \citet[][hereafter M21]{magrini21}. We refer to that paper for a description of the Gaia-ESO analysis and of the membership selection procedure.

\begin{table}[t]
\caption{Six open clusters with well populated giant branches.} 
\centering
\tiny{
\begin{tabular}{lcccccc}
\hline
\hline
Cluster &  Age & D    &  R$_{\rm GC}$ &[Fe/H] & MSTO & RC  \\
       &  (Gyr) & (kpc)     & (kpc)   & (dex)  &  (M$_{\odot}$) & (M$_{\odot}$) \\
\hline

NGC 2158    & 1.5&   4.3  & 12.6&   -0.16$\pm$0.05  & 1.7 & 1.8 \\
NGC 2141    & 1.9 &  5.2  & 13.3 & -0.06$\pm$0.07  & 1.6  & 1.7\\
Berkeley 21       & 2.1 &  6.4  &14.7 & -0.21$\pm$0.04  & 1.5 & 1.6 \\ 
Trumpler 5  & 4.3 & 3.0 &11.2 & -0.35$\pm$0.04  &1.1  & 1.2 \\
NGC 2243    & 4.4 & 3.7 &10.6 & -0.44$\pm$0.09  &1.1  & 1.2 \\
Berkeley 32       & 4.9 & 3.1 &11.1 & -0.28$\pm$0.08  &1.1  & 1.2  \\
\hline
\end{tabular}
}
\label{tab:clusters}
\tablefoot{Age, distance, and Galactocentric distance are from \citet{CG20}. The mean [Fe/H] are from the members observed with the UVES spectrograph in Gaia-ESO {\sc idr6}. The MSTO and RC masses are obtained using {\sc Parsec} isochrones \citep{bressan12}.}
\end{table}

We define two different samples: the first is composed by all cluster members  with 1~$M_{\odot}\le M_\mathrm{MSTO}\le 1.8~M_{\odot}$ and a restricted range of metallicity,
$-0.2\le$ [Fe/H] $\le+0.2$~dex, belonging to the clusters Col261, Be39, NGC6791, M67, Haf10, Cz24, NGC2425, Trumpler20, NGC2141, NGC2420, NGC2158, NGC2154; we use this sample for a first global comparison with the models. For the second sample, we select six open clusters with well populated giant branches for a more detailed comparison with the models. These clusters have 1550~Myr$\,\le\,$ age $\,\le\,$ 4900~Myr, host at least $\sim$20 red giant stars for which Li abundance is available, present a clearly distinguishable RC, and have $1.1\,M_{\odot} \leq M_{\rm MSTO} \leq\,1.7\,M_{\odot}$ (Table~\ref{tab:clusters}).

Lithium in the selected clusters was measured from UVES spectra for RC stars, and either from UVES (if available) or GIRAFFE data for the remaining members (see Fig.~\ref{fig:spectra} for two examples of spectra). When the line was too weak and barely or not visible, upper limits were provided (see M21 for details).
Lithium abundances in Gaia-ESO are derived using one-dimensional (1D) model atmospheres in local thermodynamical equilibrium (LTE). M21 estimated that the 
abundance corrections based on more realistic 3D non-LTE model atmospheres \citep{wang21} are within $\pm$0.1~dex, depending on $T_\mathrm{eff}$, and almost negligible for MSTO stars and for giant stars hotter than 4200~K. We thus adopt the 1D LTE Gaia-ESO Li abundances. 

\section{Li abundance in RC stars}
\label{sec:rc}


\begin{figure*}[ht!]
\centering
\includegraphics[width=0.98\hsize]{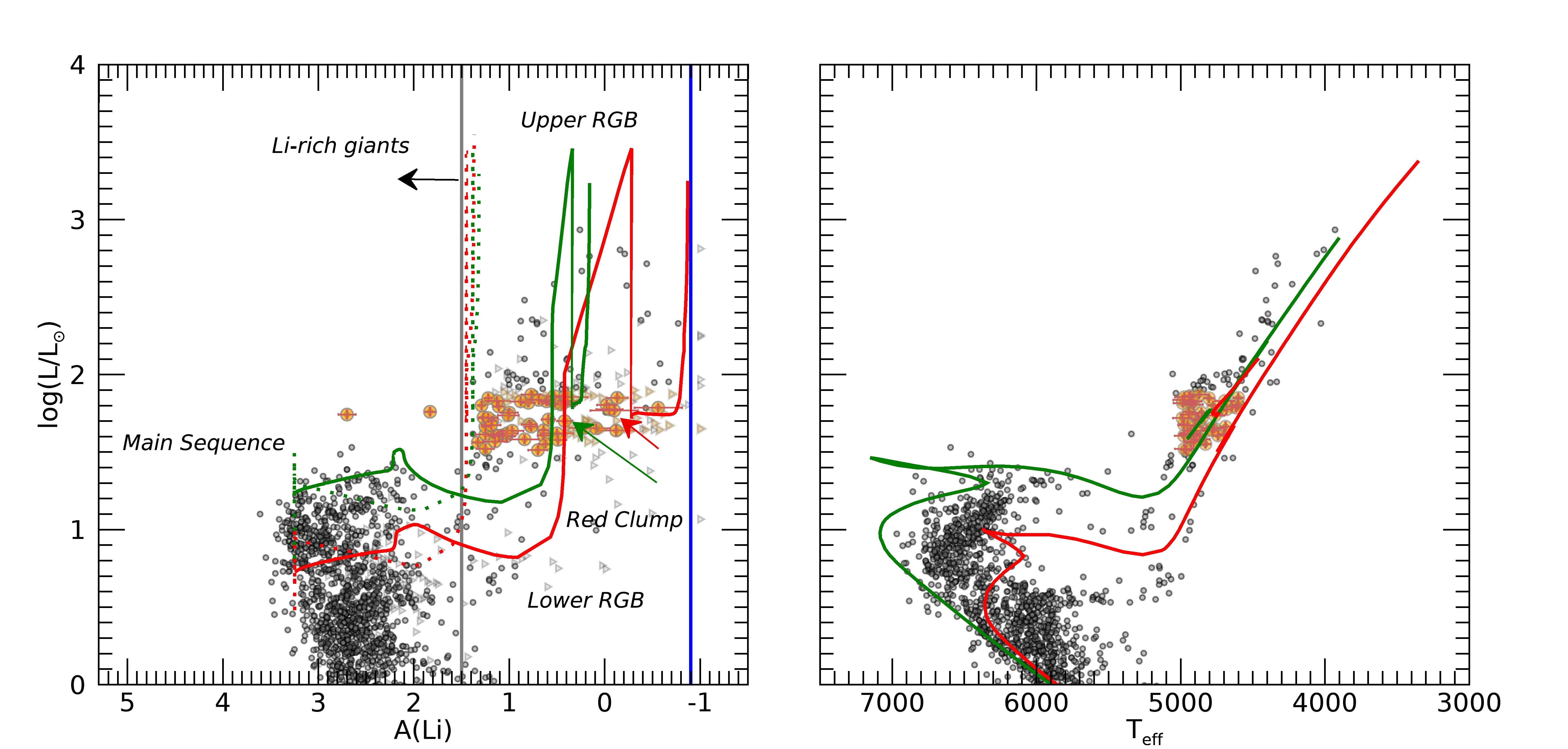}
\caption{ Left panel: $\log(L/L_{\odot})$ versus A(Li) for member stars of open clusters with $1 M_{\odot}\le M \le 1.8 M_{\odot}$ and $-0.2\le$  [Fe/H]$ \le 0.2$: grey circles show the whole sample of member stars  and coloured circles the stars at the RC. Triangles show the upper limits on A(Li).  The red and green continuous curves are the RT models at solar metallicity for 1.5~$M_{\odot}$ and 2~$M_{\odot}$, respectively, while the red and green dotted curves are the classical models for the same masses from \citet{lagarde12}. 
The vertical black line indicates the limit for Li-rich giants  A(Li)$\,\ge 1.5$~dex, while the vertical blue line at A(Li)$\,\ge -0.9$~dex shows the RC/RGB limit of \citet{kumar20b}. The locations of the RC  in the RT models of \citet{lagarde12} are shown with arrows. In the right panel, we show the HR diagram with the sample stars and the {\sc Parsec} isochrones for 1.5 and 2.0~M$_{\odot}$, in red and green, respectively, at solar metallicity.  }
\label{fig:rc}
\end{figure*}

Recently, \citet{kumar20b} investigated Li abundances in field RC stars with masses below 2.0~$M_{\odot}$, using the results of the GALactic Archaeology with HERMES (GALAH) survey \citep[DR2,][]{galahdr2}. They found the RC stars to have enhanced Li when compared to stars at the RGB tip, and with respect to a 1.0~$M_{\odot}$ model that includes effects of extra mixing (thermohaline instability and rotation-induced mixing). This led to the suggestion that Li production is a general phenomenon that affects all low-mass stars between the RGB tip and the RC.


To test the above hypothesis, we identify RC stars in our first sample of cluster members by comparing their location in the Hertzprung-Russel  diagrams  with isochrones. We chose as RC stars those with $4600\le T_\mathrm{eff} \,\mathrm{(K)} \le 5000$ and $1.5\le\log(L/L_{\odot})\le 1.9$ (see Fig.~\ref{fig:rc}).
With this selection, there might still be some contamination from RGB stars; however, the probability is low. A a star with a mass between 1 and 1.8$M_{\odot}$, during the RGB phase, spends approximately 0.2-0.4$\times$10$^8$~yr with the same luminosity of a RC star, while it stays in the RC phase 2-5 times longer \citep[see also][for an estimate of the timescales of the RC phase]{sing21}.  
In addition, the difference in $T_\mathrm{eff}$ between the two phases at the luminosity of the RC varies from $\sim$300~K at 1~$M_{\odot}$ to $\sim$150~K at 1.8~$M_{\odot}$, considerably larger than our typical errors (30-60~K) allowing us to separate the two phases. In our sample, we have 53 RC stars with A(Li) measurements and 50 with upper limits. 
The CNAMEs, cluster to which they belong, stellar parameters, lithium abundances and MSTO masses of the selected RC stars are reported in Table~\ref{table:li:rc:all}. 

\begin{table*}
\caption{Red clump member stars of open clusters with $1 M_{\odot}\le M_{\rm MSTO} \le 1.8 M_{\odot}$ and $-0.2\le$ [Fe/H]$ \le$ 0.2. The full table is available online at the CDS.}
\label{table:li:rc:all}
\tiny
\begin{tabular}{lllllllll}
\hline\hline
  \multicolumn{1}{c}{CNAME} &
  \multicolumn{1}{c}{Cluster} &
  \multicolumn{1}{c}{$T_\mathrm{eff}$} (K) &
  \multicolumn{1}{c}{log~g} &
  \multicolumn{1}{c}{[Fe/H]} &
  \multicolumn{1}{c}{A(Li)}  &
  \multicolumn{1}{c}{UL$_{\rm A(Li)} ^a$} &
  \multicolumn{1}{c}{log (L/L$_{\odot}$)} &
  \multicolumn{1}{c}{MSTO (M$_{\odot}$)} \\
\hline
 07465009-0436004 & Berkeley39    &   4820$\pm$30   & 2.62$\pm$0.05  & -0.14$\pm$0.17 & 0.64$\pm$0.05 & 0   &               1.64$\pm$ 0.03 & 1.2 \\
  07470378-0439141 & Berkeley39    &    4700$\pm$30  & 2.52$\pm$0.05  & -0.14$\pm$0.05 & 1.17$\pm$0.06 &  0  &                1.62$\pm$0.03 & 1.2 \\
\hline\end{tabular}
\tablefoot{$^a$ upper limits are indicated with 1, detections with 0}
\end{table*}


In Fig.~\ref{fig:rc}, we plot the stars in the  A(Li)-$\log(L/L_{\odot})$ plane. Two sets of  models at solar metallicity for  1.5 and  2~$M_{\odot}$ stars from \citet{lagarde12} are shown. One set includes only effects of mixing due to convection. The second includes, in addition, rotation-induced mixing and thermohaline instability (hereafter, RT models). The figure shows the evolution of the lithium abundances: on the left, we have stars at the end of the main sequence, where A(Li) is between 2.5 and 3.4~dex. Afterwards, there is a first episode of dilution at the first dredge-up (FDU) which, in classical models, gives a value of A(Li) $\sim$ 1.3-1.5, almost independent of stellar mass. In the RT models, the FDU results in lower Li, because of the effects of rotation during the main sequence. In addition, there is a further dilution of Li at the RGB bump when the thermohaline instability is activated. After that, stars evolve towards the RGB tip  and then drop in luminosity, reaching the RC at $\log (L/L_{\odot})\sim$1.5-1.8. The models suggest Li abundances as low as $-0.3$~dex. Further evolution during the clump can deplete Li down to $-$1~dex, before the stellar luminosity increases again. 

The RC stars in our sample with $1 M_{\odot}\le M< 1.8  M_{\odot}$ have $-0.9\le$ A(Li) $\le 1.3$. This is higher than predicted by RT models but lower than predicted by classical models (excluding the two  Li-rich giants). The mean A(Li) of the RC stars is A(Li) $=0.78\pm 0.55$~dex (only measurements, no upper limits). This value is close to the peak of the distribution obtained by \citet{kumar20b}, A(Li)$\sim$0.7~dex. These data are thus consistent with their findings and suggest that there might be a further Li enrichment during the RC phase. 

\subsection{Li in RC of individual clusters}\label{sec:rc.clusters}

\begin{figure*} 
\centering
\includegraphics[scale=1.0]{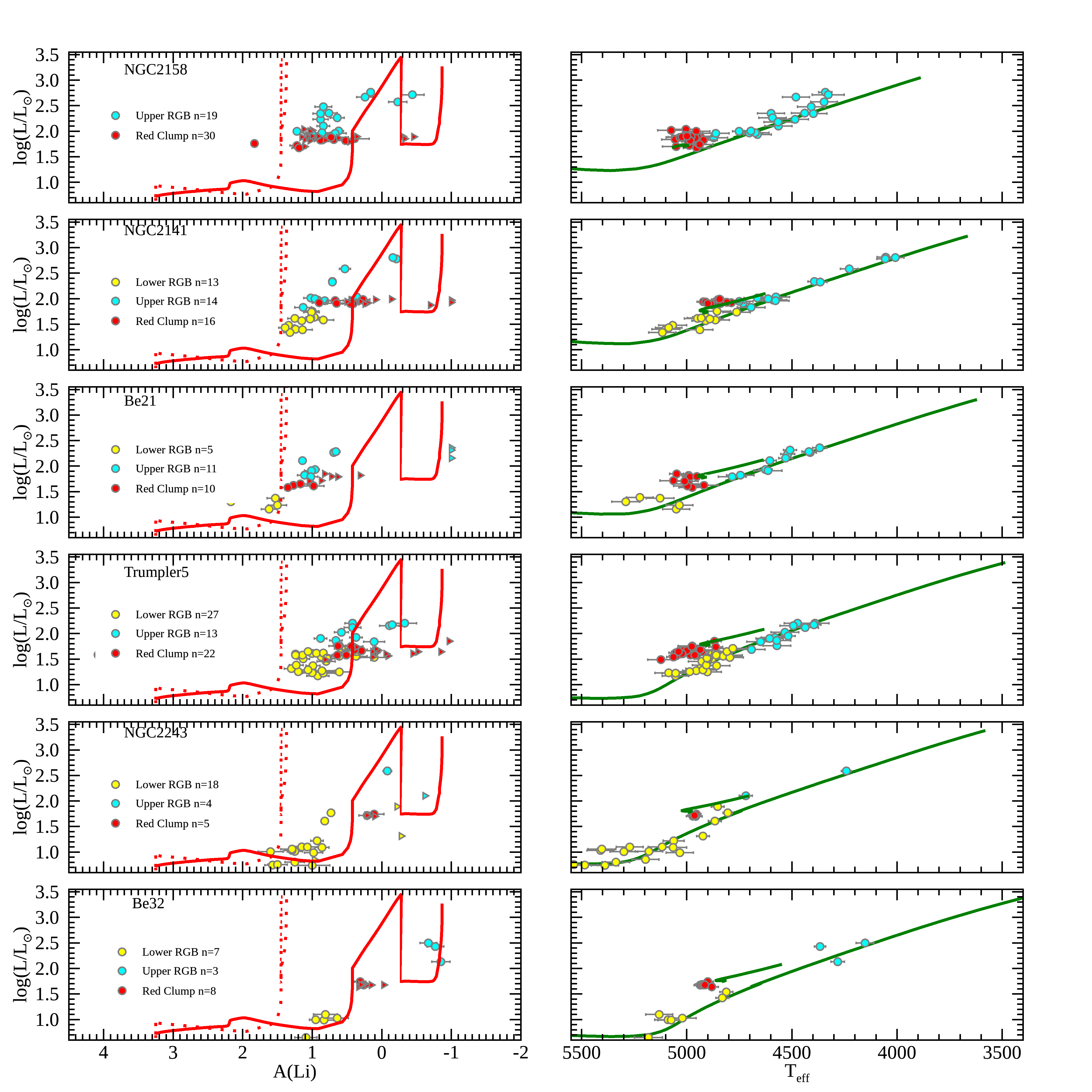}
\caption{Left panels: $\log(L/L_{\odot})$ versus A(Li) in our sample clusters: yellow circles are lower RGB stars -- prior to the RGB bump --,  cyan circles are upper RGB stars -- after the bump -- and red circles are RC stars. The  red continuous curves are the models with rotation-induced mixing at solar metallicity for 1.5~$M_{\odot}$. The red dotted curves are the classical models with only convection from \citet{lagarde12}. 
Right panels:  HR diagrams with the {\sc Parsec} isochrones \citep{bressan12} at the age and metallicity of each cluster. 
The error bars on luminosity are of the order of the symbol size}.
\label{fig:all}
\end{figure*}

Figure~\ref{fig:all} shows, separately, the six clusters of Table \ref{tab:clusters} in the $\log(L/L_{\odot})$ versus A(Li) plane and in the HR diagram. 
The CNAMEs, cluster to which they belong, stellar parameters, lithium abundances, MSTO mass and evolutionary phase of the selected RGB and RC stars are reported in Table~\ref{table:li:clusters}. 

\begin{table*}
\caption{RGB and RC member stars of the six open clusters of Table~\ref{tab:clusters}. The full table is available online at the CDS.}
\label{table:li:clusters}
\tiny
\begin{tabular}{llllllllll}
\hline\hline
  \multicolumn{1}{c}{CNAME} &
  \multicolumn{1}{c}{Cluster} &
  \multicolumn{1}{c}{$T_\mathrm{eff}$} (K) &
  \multicolumn{1}{c}{log~g} &
  \multicolumn{1}{c}{[Fe/H]} &
  \multicolumn{1}{c}{A(Li)}  &
  \multicolumn{1}{c}{UL$_{\rm A(Li)} ^a$} &
  \multicolumn{1}{c}{log (L/L$_{\odot}$)} &
  \multicolumn{1}{c}{MSTO (M$_{\odot}$)} &
  \multicolumn{1}{c}{Phase}\\
\hline
 05513791+2143345  & Berkeley21 &        5223$\pm$64   &  3.63$\pm$0.18 &    -0.34$\pm$0.06 &    0.83                  &    1            &        1.39$\pm$0.03    &    1.5 &   LRGB   \\
  05515065+2148321  & Berkeley21 &        5127$\pm$66    & 3.21$\pm$0.18  &   -0.07$\pm$0.05 &   1.53$\pm$0.12  &  0              &      1.37$\pm$0.03     &    1.5  &  LRGB   \\
\hline\end{tabular}
\tablefoot{$^a$ upper limits are indicated with 1, detections with 0}
\end{table*}

As discussed in M21, for low-mass solar-type stars with relatively extended convective envelopes,  hydrodynamic processes induced by rotation, such as meridional circulation and shear mixing, predict large rotation gradients within the interior, which for instance require internal gravity waves or other  mechanisms, such as penetrative convection, tachocline mixing, and additional turbulence to explain the rotation profile and the surface abundance of lithium in solar-type stars of various ages. 
These additional mechanisms are not included in the RT models of \citet{lagarde12} for 1 M$_{\odot}$ stars. 
Since in RT models for 1 M$_{\odot}$ stars at solar metallicity, the depletion of Li in the RGB phase reaches extremely low values  not corresponding to the observed abundances, we consider  
more suitable  to compare the observations of all our sample clusters with the models for 1.5~M$_{\odot}$.

The stars are divided in lower RGB (if below the luminosity bump when compared to an isochrone), upper RGB (if above the bump), and RC. 
It is clear that the distributions of Li abundances in the RC stars of each cluster show values that are either similar or even higher than those of the upper RGB stars.
In standard models, the Li abundance reaches $\sim$ 1.5 dex after the first dredge-up with no further changes up to the RC. Traditionally, only giants with A(Li) $>$ 1.5 are considered to be Li-rich. In the RT model, rotation-induced mixing brings A(Li) down to $\sim$ 0.4 after the first dredge-up and thermohaline mixing reduces it to $\sim$ $-$0.3 dex after the bump.
Figure \ref{fig:all} shows that each evolutionary stage, in each cluster, is characterised by a spread in Li abundances. This can be explained by a spread in the initial stellar rotation and thus in the effects of the rotation-induced mixing \citep[e.g.][]{c20}. In other words, that a given star, in any evolutionary stage, has an observed Li abundance above the model prediction does not immediately imply that Li was produced. Observed values above the model prediction can be explained by a weaker action of rotation-induced mixing and initial rotation is a property known to vary considerably from star to star \citep[e.g.][]{2013A&A...556A..36G}.

To determine whether or not there is Li production in the RC, one has to investigate the general properties of the abundance distributions in each evolutionary stage. The differences in the distribution of Li abundances between lower RGB and RC stars indicates a certain level of Li depletion by thermohaline mixing. The brighter RGB stars in NGC 2158 and NGC 2141 show signs of extra Li depletion after the bump. It seems fair to assume that the depletion will increase before the stars reach the RGB tip and the RC. Therefore, it indeed seems that fresh Li production is needed to explain how the RC stars can have similar or higher Li than the upper RGB stars.

This agrees with the conclusions of \citet{kumar20b}. However, we remark that about 65\% of our RC stars  have Li upper limits, some at the same level of the detected abundances. The remaining 35\% RC stars have Li measurements, at the level of the lower RGB abundances and higher than the upper RGB ones. Thus strong Li depletion in some of the giants cannot be excluded. Either the Li-rich stage is short-lived \citep[see][]{sing21} or Li production is not ubiquitous. In the GALAH data used by \citet{kumar20b}, upper limits are not flagged. If some of the Li abundances they discuss are actually upper limits, then perhaps not all of their RC stars have high Li abundance.

\section{Possible mechanisms to explain Li in RC stars}

\begin{figure*}[ht!]
\centering
\includegraphics[scale=0.77]{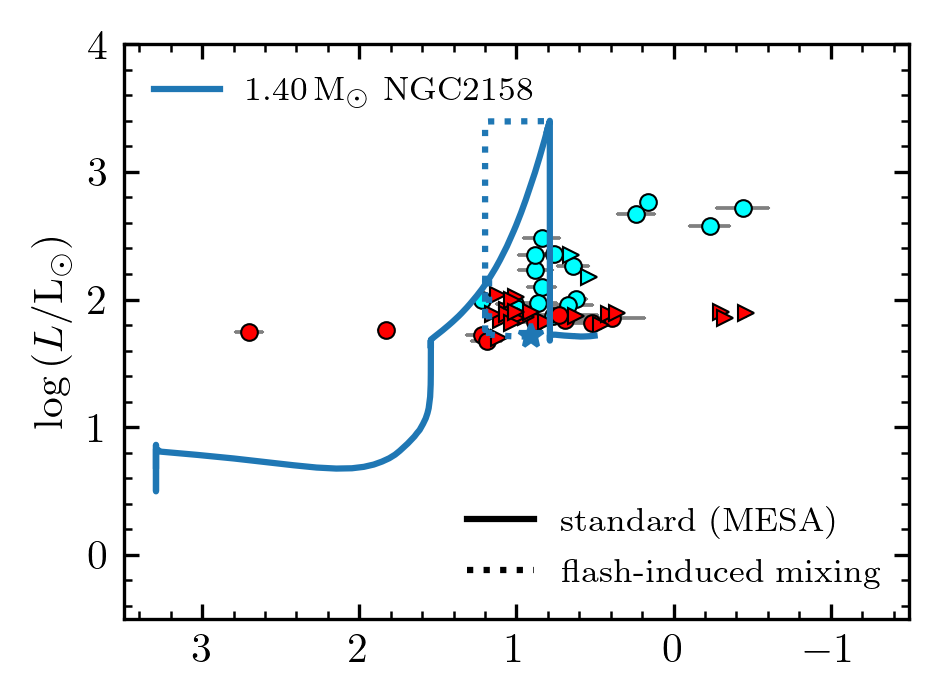}\includegraphics[scale=0.77]{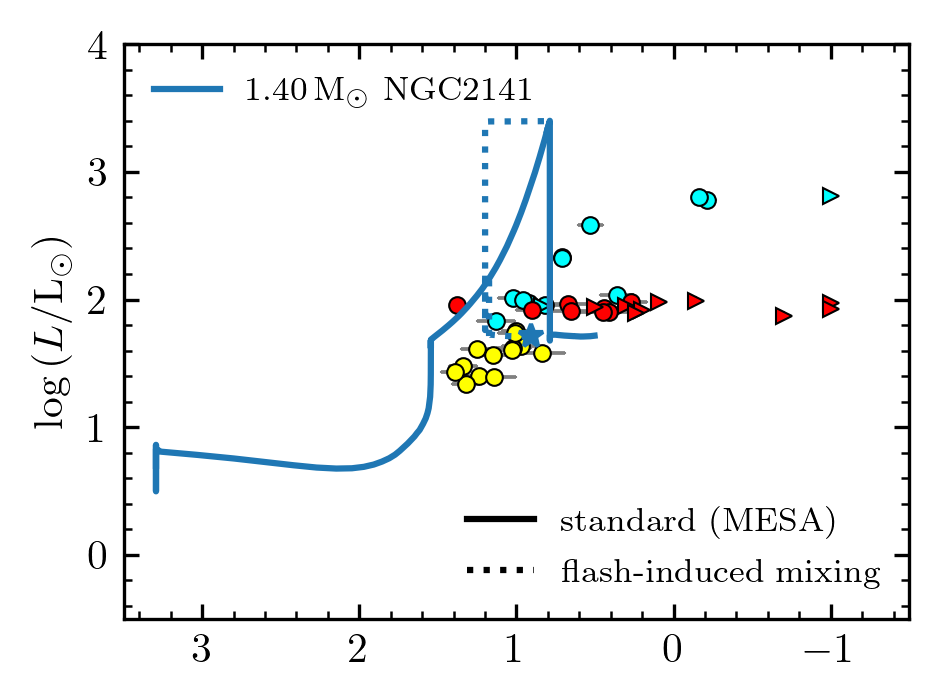}\includegraphics[scale=0.77]{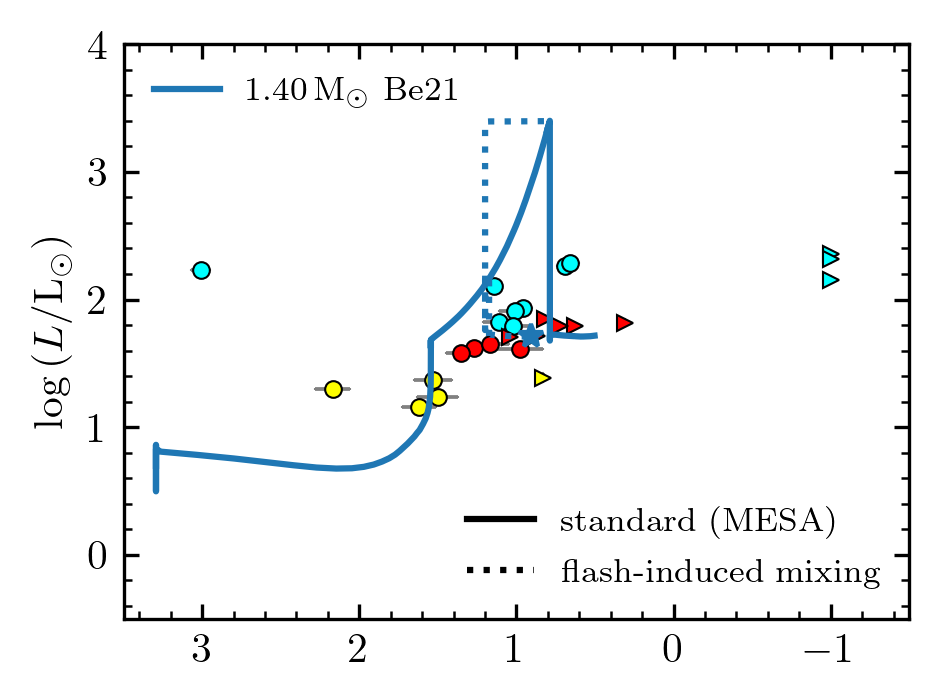}
\includegraphics[scale=0.77]{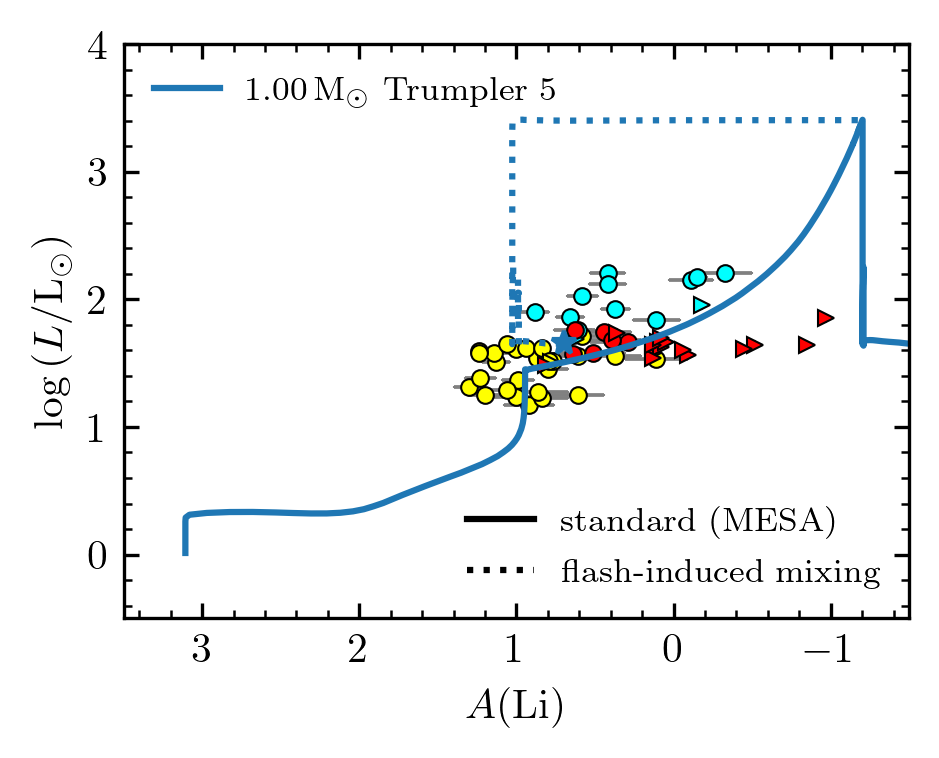}\includegraphics[scale=0.77]{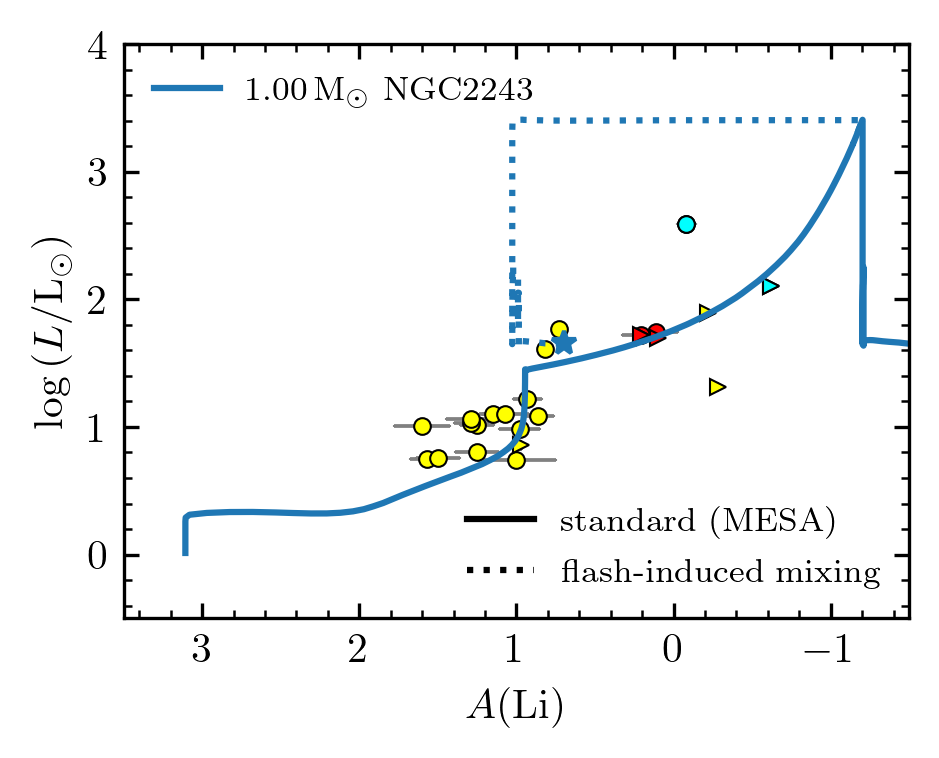}\includegraphics[scale=0.77]{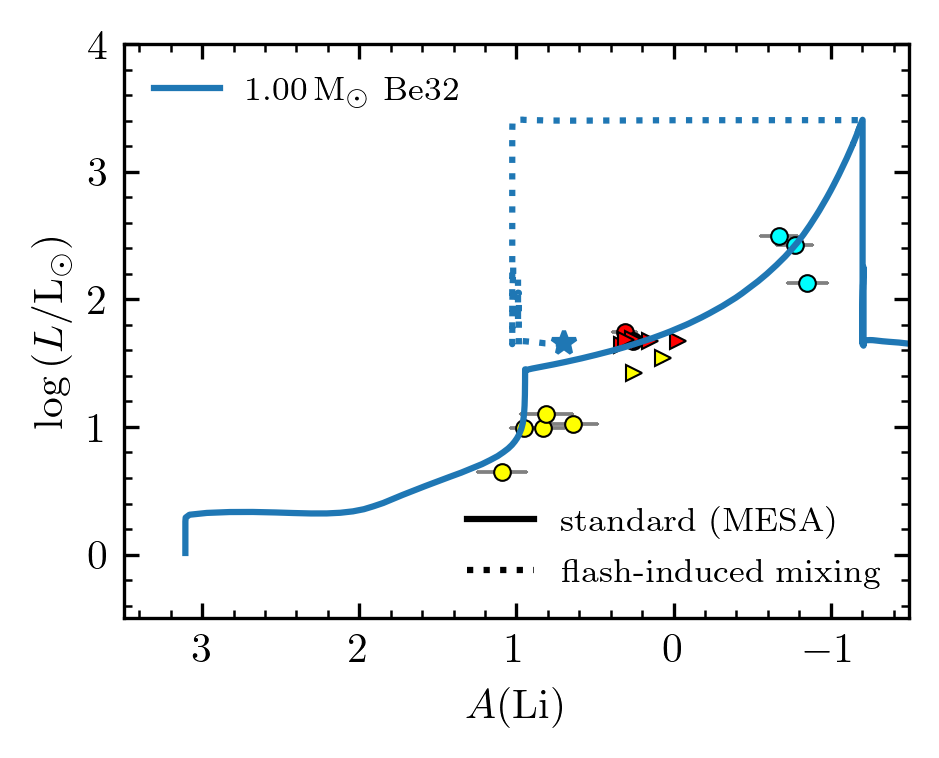}
\caption{Luminosity versus A(Li) in NGC~2158, NGC~2141, Be~21, Trumpler~5, NGC~2243 and Be~32. Symbols are the same as in Fig.~\ref{fig:all}. The blue continuous curves are the {\sc MESA} models with thermohaline mixing. The blue dotted curves are the models with flash-induced mixing of \citet{schwab20}. The stellar mass of the models is indicated in each panel.}
\label{fig:flash}
\end{figure*}

\subsection{Mixing induced by the first He flash}


Motivated by the Li discrepancy highlighted by \citet{kumar20b} between tip-RGB and RC stars, \citet{schwab20} recently proposed a model where the He flash is connected to the surface Li enhancement. The model of \citet{schwab20} is constructed using Modules for Experiments in Stellar Astrophysics \citep[{\sc MESA}, see e.g.][]{paxton19}. Thermohaline instability is included but rotation-induced mixing is not taken into account, being thus different from the RT models. Lithium production is a result of the Cameron-Fowler (CF) process \citep{cf71}, out of the decay of $^{7}$Be produced in internal regions with temperatures above $\sim$10$^{7}$ K. In their model, the mixing in the envelope is triggered by the first and strongest He sub-flash. The physical mechanism that induces this mixing might be due to internal gravity waves \citep[see][for a discussion]{schwab20}.


In Fig.~\ref{fig:flash} we compare A(Li) in our sample of six clusters with the results of models with and without the flash-induced mixing from  \citet{schwab20}\footnote{\url{https://doi.org/10.5281/zenodo.4688026}}. The effect of thermohaline-induced mixing and of the flash-induced mixing both depend on mass and are stronger for lower masses. For the three younger clusters (NGC~2158, NGC~2141 and Be~21), the models without the flash-induced mixing for 1.4~$M_{\odot}$ can reproduce quite well the A(Li) in the RC stars. The extra mixing during the He flash is not totally necessary to explain A(Li) in the RC. The results are similar to those obtained with the RT models presented in Fig.~\ref{fig:all}.  We note, however, that upper RGB stars have lower A(Li) than predicted by the models of \citet{schwab20}, probably because the effect of rotation is not included. However, our aim is to make a qualitative comparison, and introducing the effect of rotation goes beyond what we set out to do in this paper.  For the oldest clusters (Trumpler~5, NGC~2243, Be~32), A(Li) in RC stars is  definitively higher than the prediction of models without flash-induced mixing. The agreement is better with models that include the new extra-mixing after the tip of the RGB. Nevertheless, our data seem to indicate a slightly lower efficiency of the diffusive mixing process than the one adopted in \citet{schwab20}.  

\subsection{Mixing induced by neutrino magnetic moment}


To explain the ubiquitous enhancement of Li in RC stars, \citet{mori21} introduced an additional energy loss related to the neutrino magnetic moment (NMM), $\mu_{\nu}$ \citep[see also][]{mori20}. They used the {\sc MESA} code to build a fiducial model with $\mu_{\nu}$ = 0 and a set of models with $\mu_{12}$ ranging from 1 to 5, where $\mu_{12}=\mu_{\nu}$/10$^{-12}\mu_{B}$ and $\mu_{B}$ is the Bohr magneton. Assuming $\mu_{\nu}>0$, the He flash is delayed and the CF mechanism can continue to produce Li. The delayed He flash results in stars with heavier He core and increased luminosity at the RGB tip. More massive cores have a smaller density above the hydrogen burning shell and, consequently, a larger thermal conductivity \citep{lattanzio15} and a more effective thermohaline mixing. The enhanced mixing increases the amount of $^7$Be transported to the surface convective layer, resulting in higher Li in RC stars.

\begin{figure*}[ht!]
\centering
\includegraphics[scale=0.77]{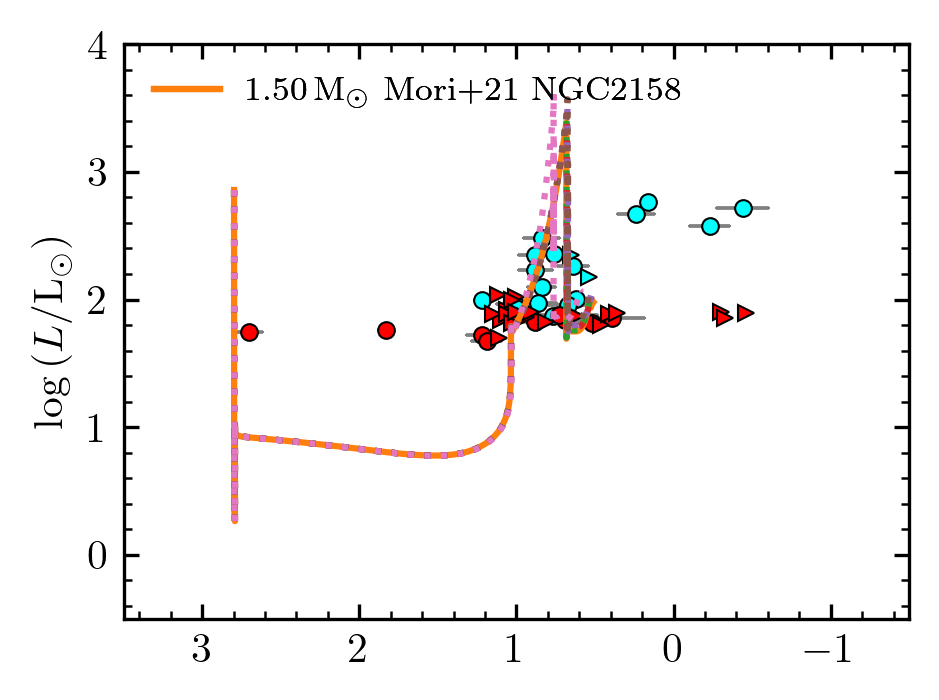}\includegraphics[scale=0.77]{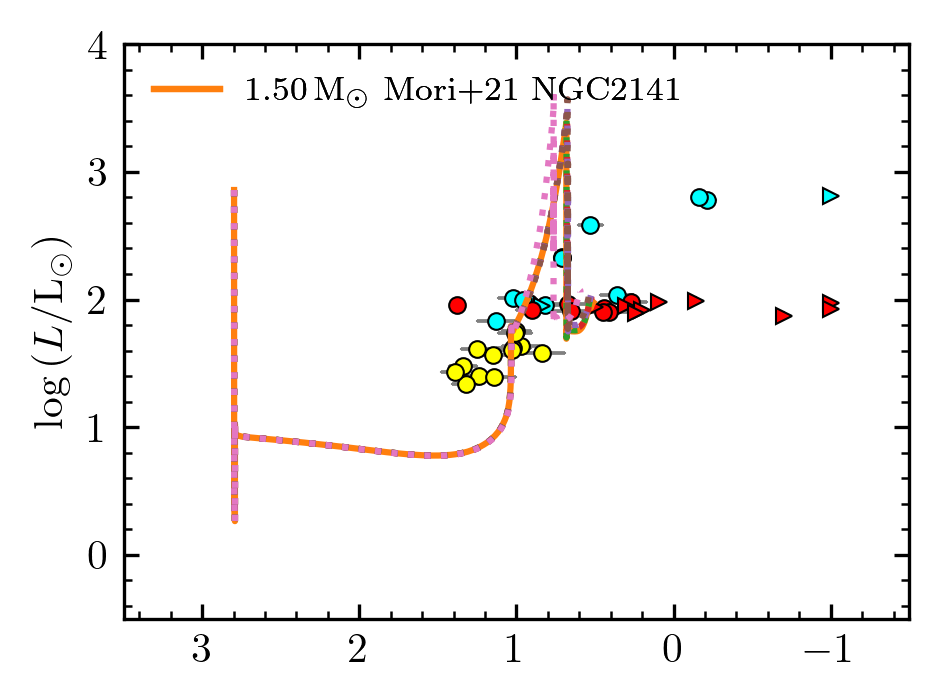}\includegraphics[scale=0.77]{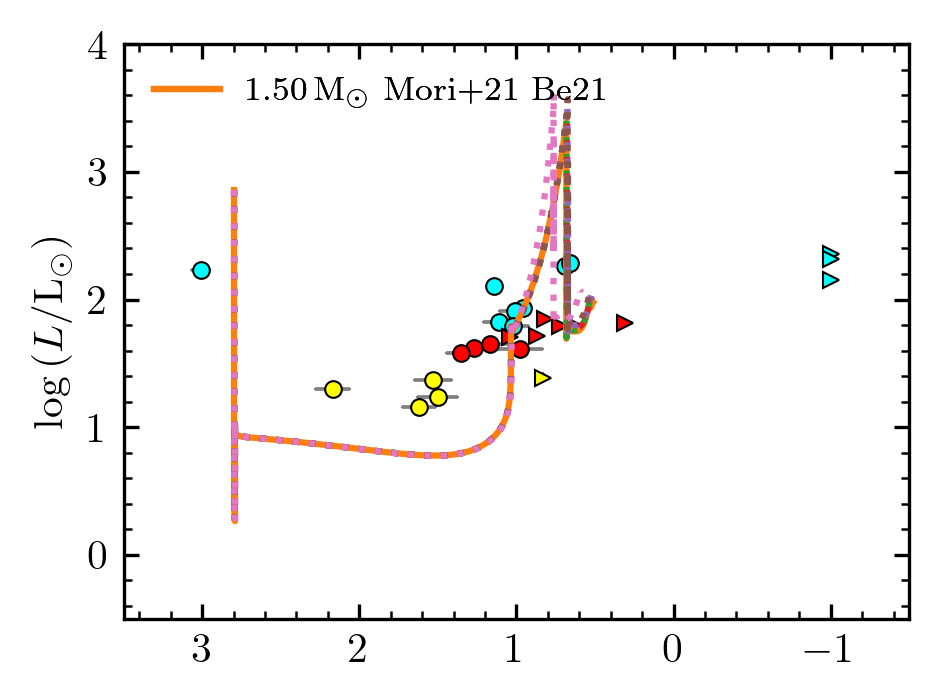}
\includegraphics[scale=0.77]{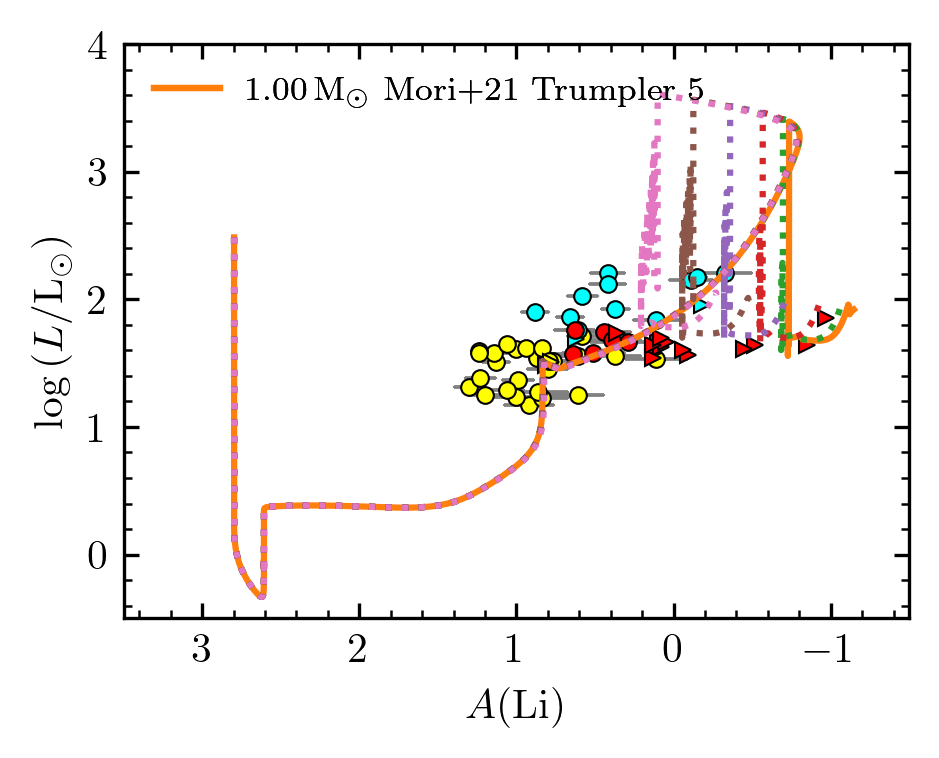}\includegraphics[scale=0.77]{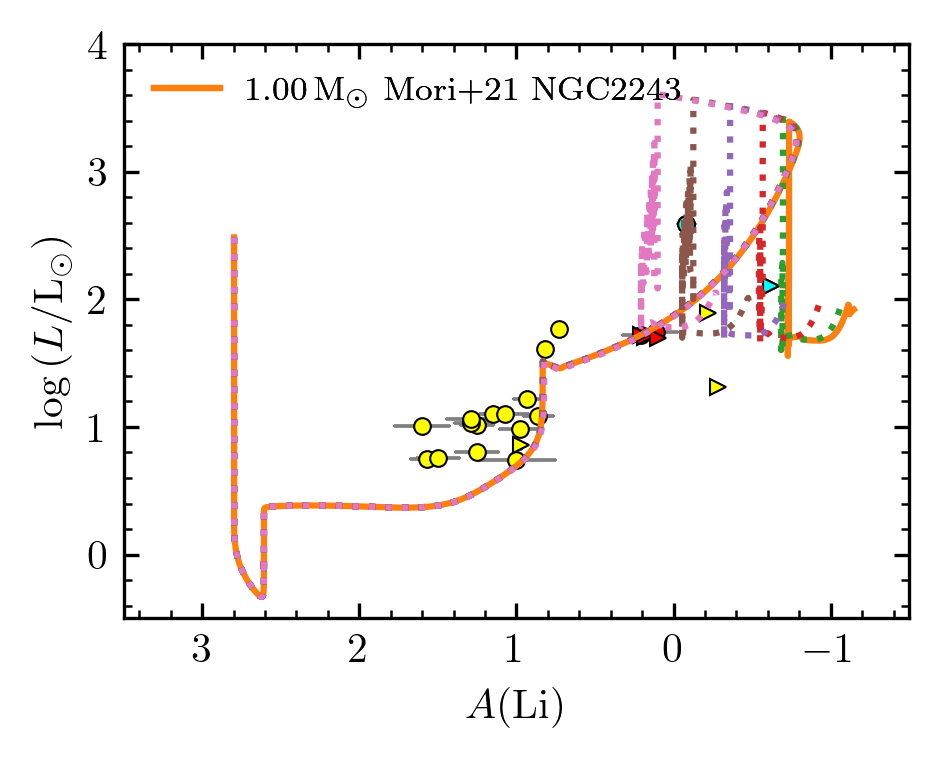}\includegraphics[scale=0.77]{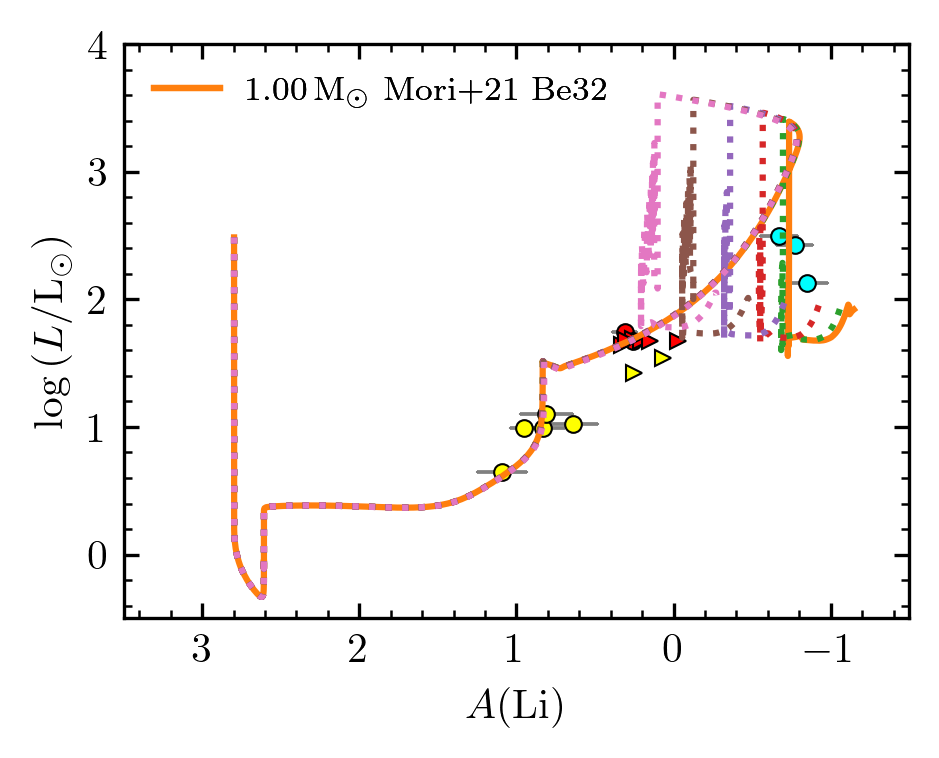}
\caption{$\log(L/L_{\odot})$ versus A(Li) in NGC~2158, NGC2141, Be~21, Trumpler~5, NGC~2243 and Be~32: symbols and colours for the observations as in Fig.~\ref{fig:flash}.  The orange continuous curves are the {\sc MESA} models with thermohaline mixing for the masses indicated in each panel, and $\mu_{12}$=0. The  dotted curves are the models with NMM mixing of \citet{mori21} for 1~M$_{\odot}$ and different $\mu_{12}$, from 1 to 5, from right to left.   }
\label{fig:NNM}
\end{figure*}

In Fig.~\ref{fig:NNM}, we compare our observations with the \citet{mori21} models. The effect of NMM for the younger clusters is limited, and there are negligible differences with respect to the standard models. 
As in the  models of \citet{schwab20}, the standard model is not able to reproduce the depletion observed towards the RGB tip. 
For the three older clusters, the standard models would predict lower A(Li) in the RC stars, while  models with higher $\mu_{12}$ cover quite well the observed range of A(Li).  
However the highest values of NNM are slightly higher than current astrophysical limits for this quantity, so also other
channels may play a role to explain the behaviour of RC stars. 

Both the models of \citet{schwab20} and \citet{mori21} seem able to explain the observations of A(Li) in low-mass RC stars ($M\sim 1~M_{\odot}$). The two scenarios, however, differ in the time-scales of the processes, as described in \citet{mori21}. In the He-flash mixing model, A(Li) is enhanced during the helium flash. In the NMM model, the enhancement happens on a longer time scale, i.e. about 1~Myr before the flash. As suggested by \citet{mori21}, it might be possible to distinguish between the two models by searching for lithium-rich giant stars with A(Li)$\sim$0 near the tip of the RGB.

We briefly remark that \citet{masseron17} identified significant N depletion in field stars between the RGB tip and the RC, supporting the need of extra mixing between these phases. In this case, nitrogen has to be burnt near or at the He flash episode before the material is mixed to the surface.

\section{Summary and conclusions}
\label{section:SummaryConclusions}

We investigate the evolution of A(Li) from the MSTO to the RGB and RC phases using a sample of giants in open clusters from Gaia-ESO {\sc idr6}. We find RC stars where Li is detected at a level 
consistent with or above those of upper RGB stars. This suggests that there might be some Li enrichment between these phases which results in stars that would normally not be classified as Li-rich. It would be useful to determine the carbon isotopic ratio in these RC stars, and confirm that they have gone through the extra mixing after the luminosity bump.

Nevertheless, our sample has several RC giants with Li upper limits, across the whole mass range. These limits are, for about 65\% of our RC stars in individual clusters, at the same level of the detected Li abundances (the lines are always very weak and hard to detect, see Appendix \ref{ap:spectra}.). Upper limits could be hiding strong Li depletion in these stars. It seems that upper limits are not properly flagged in the data used by \citet{kumar20b}. Therefore, their sample might also include RC stars with strong Li depletion.  However, for 35\% of our RC stars in the clusters of Table~\ref{tab:clusters}, the Li abundance is at the level of that in lower RGB stars, and usually higher than in upper RGB stars. Given these percentages, we cannot conclude that Li enrichment is ubiquitous in RC stars, as their Li abundances could be much lower, but that it might happen in a large percentage of them: from at least one third  up to one half of them (see the sample in Fig.~\ref{fig:rc}) .

The comparison with models that include additional mixing processes, such as, e.g., the He-flash mixing \citep{schwab20} and the NMM mixing \citep{mori21}, is very promising and can qualitatively explain the behaviour of the RC stars. Both models work quite well to explain the behaviour of low-mass stars ($M\sim\,1\, M_{\odot}$). They differ in the description of the processes that activate the CF mechanisms and on their time scale. They agree on the requirement of a process of mixing during the He flash, needed to activate the production of Li.

\begin{acknowledgements}
We thank an anonymous referee for her/his careful reading of the manuscript and for useful and constructive comments. 
We thank J. Schwab for recomputing for us his models (available at https://doi.org/10.5281/zenodo.4688026), and for useful discussion and comments. We thank K. Mori for kindly providing us his models to compare with our data. 
Based on data products from observations made with ESO Telescopes at the La Silla Paranal Observatory under programme ID 188.B-3002. These data products have been processed by the Cambridge Astronomy Survey Unit (CASU) at the Institute of Astronomy, University of Cambridge, and by the FLAMES/UVES reduction team at INAF/Osservatorio Astrofisico di Arcetri. These data have been obtained from the Gaia-ESO Survey Data Archive, prepared and hosted by the Wide Field Astronomy Unit, Institute for Astronomy, University of Edinburgh, which is funded by the UK Science and Technology Facilities Council.
This work was partly supported by the European Union FP7 programme through ERC grant number 320360 and by the Leverhulme Trust through grant RPG-2012-541. We acknowledge the support from INAF and Ministero dell' Istruzione, dell' Universit\`a' e della Ricerca (MIUR) in the form of the grant "Premiale VLT 2012". The results presented here benefit from discussions held during the Gaia-ESO workshops and conferences supported by the ESF (European Science Foundation) through the GREAT Research Network Programme.
LM, GC, AB, MVdS, EP acknowledge the funding from MIUR Premiale 2016: MITiC. MVdS and LM thanks the WEAVE-Italia consortium. LM acknowledges the funding from the INAF PRIN-SKA 2017 program 1.05.01.88.04. CVV, LM, EP thank the COST Action CA18104: MW-Gaia. LS acknowledges financial support from the Australian Research Council (discovery Project 170100521) and from the Australian Research Council Centre of Excellence for All Sky Astrophysics in 3 Dimensions (ASTRO 3D), through project number CE170100013. F.J.E. acknowledges financial support from the Spanish MINECO/FEDER through the grant AYA2017-84089 and MDM-2017-0737 at Centro de Astrobiología (CSIC-INTA), Unidad de Excelencia María de Maeztu, and from the European Union’s Horizon 2020 research and innovation programme under Grant Agreement no. 824064 through the ESCAPE - The European Science Cluster of Astronomy \& Particle Physics ESFRI Research Infrastructures project.
TB was funded by grant No. 2018-04857 from The Swedish Research Council.
\end{acknowledgements}

\bibliographystyle{aa}
\bibliography{Bibliography}

\begin{appendix}
\section{Detection of weak Li lines}\label{ap:spectra}

\begin{figure}[!ht]
\centering
\includegraphics[scale=0.60]{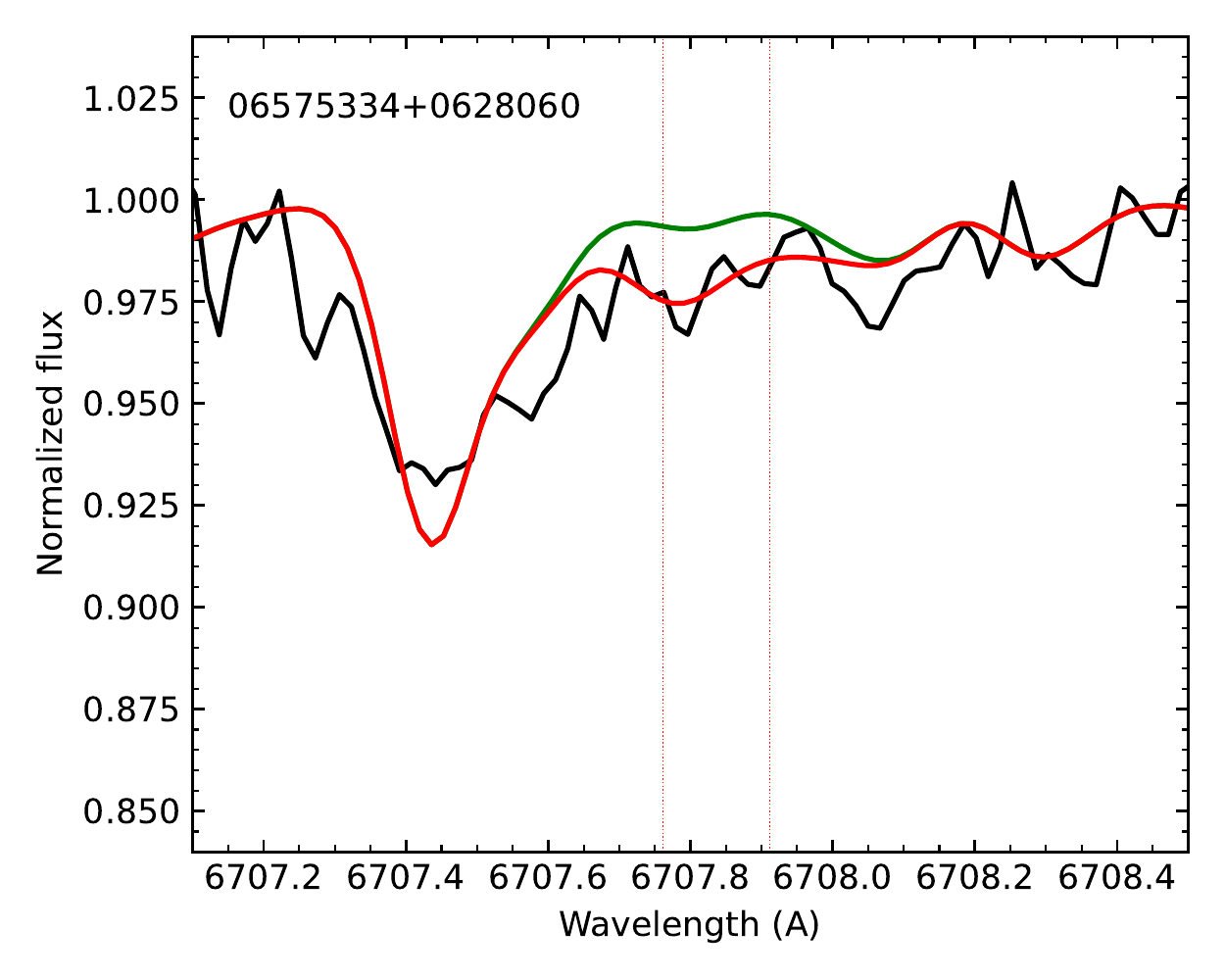}
\includegraphics[scale=0.60]{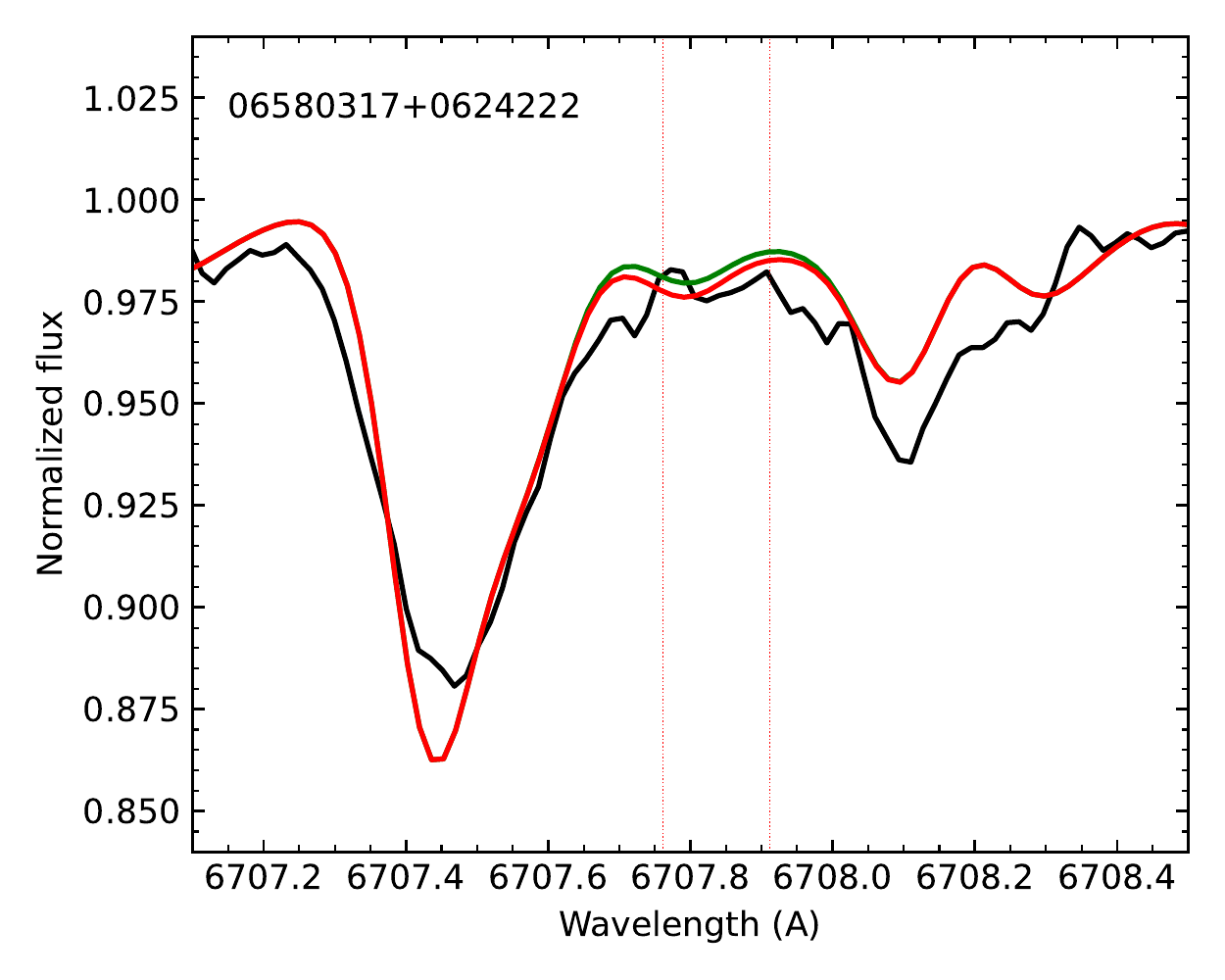}
\caption{Spectral region around the Li doublet for an RC star (upper panel) and an upper RGB star (lower panel), both members of Berkeley~32. The observed spectra are in black. The red lines show the synthetic spectra at the corresponding measured stellar parameters and A(Li) ($+0.3$ and $-0.8$~dex, respectively); the green lines show the synthetic spectra for the same parameters and A(Li)$=-$1.0. The vertical lines indicate the location of the two Li lines.  }
\label{fig:spectra}
\end{figure}

In Fig.~\ref{fig:spectra} we show two examples of the Li lines in a red clump star and in an upper RGB one observed with the UVES spectrograph. The observed spectra are compared with two sets of synthetic spectra, computed for the corresponding set of stellar parameters, but with different Li abundances, the measured one and A(Li)$=-$1.0. Although the observed Li lines are very weak, the figure clearly shows that a non-negligible amount of Li is present in the RC star, while the RGB one is compatible with A(Li)$=-$1.0. 

\end{appendix}
\end{document}